\documentclass[%
 reprint,
 amsmath,amssymb,
 aps,prx
]{revtex4-2}

\usepackage{graphicx}
\usepackage{dcolumn}
\usepackage{bm}
\usepackage{appendix}
\usepackage{xcolor}

\definecolor{darkgreen}{rgb}{0.0, 0.5, 0.0}

\begin{document}

\preprint{APS/123-QED}

\title{{Towards a Theory for the Formation of Chimera Patterns in Complex Networks}}

\author{Malbor Asllani}
 \affiliation{Department of Mathematics, Florida State University,
1017 Academic Way, Tallahassee, FL 32306, United States of America}
\email{masllani@fsu.edu}
\author{Alex Arenas}%
\affiliation{%
 Departament d’Enginyeria Informàtica i Matemàtiques, Universitat Rovira i Virgili, 43007 Tarragona, Spain
}%
\affiliation
{{Pacific Northwest National Laboratory, 902 Battelle Blvd, Richland, WA, 99354, USA}}
\email{alexandre.arenas@urv.cat}

\date{\today}

\begin{abstract}
Chimera states, marked by the coexistence of order and disorder in systems of coupled oscillators, have captivated researchers {with their existence and intricate patterns}. Despite ongoing advances, a fully understanding of the genesis of chimera states remains challenging. This work formalizes a systematic method by evoking pattern formation theory to explain the emergence of chimera states {in complex networks, in a similar way to how Turing patterns are produced}. Employing {linear stability} analysis and the spectral properties of complex networks, {we show that the randomness of network topology, as reflected in the localization of the graph Laplacian eigenvectors, determines the emergence of chimera patterns, underscoring the critical role of network structure.} In particular, this approach explains how amplitude and phase chimeras arise separately and explores whether phase chimeras can be chaotic or not. Our findings suggest that chimeras result from the interplay between local and global dynamics at different time scales. Validated through simulations and empirical network analyses, our method enriches the understanding of coupled oscillator dynamics.
\end{abstract}

\maketitle


{\section{Introduction}}
\label{sec:Intro}

Complex networks are crucial for examining complex systems in both natural and synthetic settings \cite{newman2010networks}. They are particularly instrumental to understand how different interconnected components within a system interact to manifest collective behaviors.
{One of the most fascinating collective phenomena observed in complex networks is the synchronization of oscillators, where individual components, despite their diverse functions and environments, achieve coherent oscillations} \cite{kuramoto,strogatz2004sync, pikovskij_synchronization:_2007, arenas_synchronization_2008}. Synchronization occurs in many natural and engineered settings, from neurons in the brain coordinating actions \cite{izhikevich_dynamical_2007, sporns_networks_2011} and fireflies synchronously flashing for mating \cite{john_synchronous_1976}, to power grids where it ensures stability \cite{motter_spontaneous_2013}, and communication networks where it prevents data loss and enhances efficiency \cite{Sivrikaya_Yener}. 
In the realm of synchronization, the concept of chimera states—where coherent and incoherent oscillators coexist—represents a significant theoretical challenge and has been a focus of extensive research \cite{chimera, Abrams_Strogatz, Panaggio_2015, amp_chimera, semenova2016coherence, kemeth2016classification, panaggio2016chimera, Chouzouris, maistrenko2017smallest, thoubaan2018existence, omel2018mathematics, zakharova_chimera_2020, parastesh2021chimeras}. These states illustrate a striking blend of order and disorder within networked systems{, where order is understood as amplitude coherence, or both amplitude and phase coherence. Chimera states have been found to be stable in infinite-size networks \cite{omelchenko_coherenceincoherence_2013} and transient in finite ones \cite{wolfrum_chimera_2011}. Nevertheless, rigorous results have been obtained in weak chimeras, defined as a type of invariant set exhibiting partial frequency synchronization \cite{ashwin2015weak, bick2016chaotic}.}  Despite numerous studies, complete explanations for their stability and emergence are still sought~\cite{Panaggio_2015}. Experimental studies have definitively confirmed the relevance of chimera states in various human-made setups: chemical oscillators using the Belousov–Zhabotinsky reaction with light feedback \cite{tinsley_chimera_2012, nkomo_chimera_2013}, optical systems with spatial light modulators \cite{hagerstrom_experimental_2012}, mechanically coupled metronomes on swings \cite{martens_chimera_2013}, and photoelectrochemical experiments modeling silicon oxidation \cite{schmidt_coexistence_2014}. {Furthermore, observations in both neuroscience and ecology suggest their occurrence in natural settings:} in the human brain, chimera states manifest as patterns of partial synchrony among brain regions fundamental to cognitive organization \cite{chimera_brain}, while in ecology, video recordings of \textit{Photuris frontalis} fireflies reveal spontaneous, stable chimera states where groups within a swarm flash synchronously but with a constant delay \cite{chimera_firefly}.\\
\begin{figure*}
    \centering
    \includegraphics[width=\textwidth]{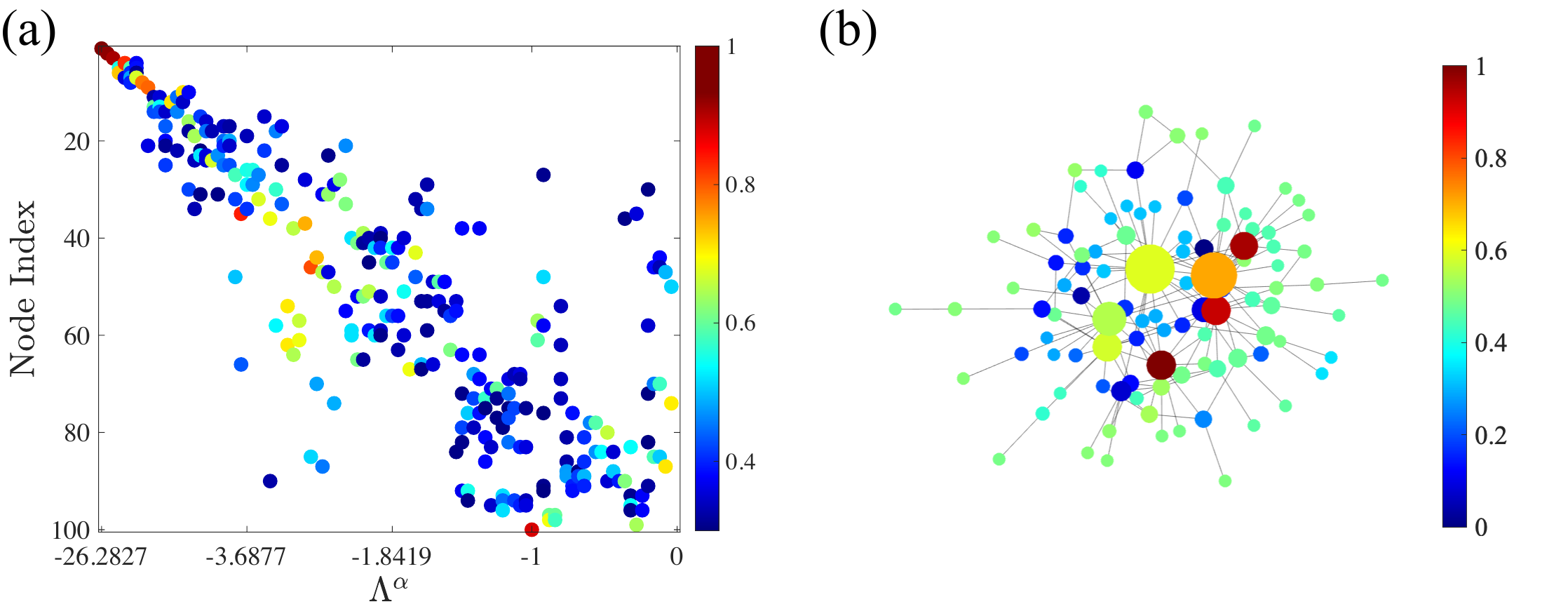}
    \caption{\textbf{Eigenvector localization and chimera states in a scale-free network.} (a) The panel displays the matrix's eigenvectors as columns, with entries below 0.3 in magnitude omitted for clarity. (b) The normalized colormap in the network visualization provides a snapshot that illustrates the amplitude of the chimera {for the Rössler model}. Referring to the color code, it can be observed that peripheral nodes (those less connected) share a similar amplitude, while the disorder is localized in the central nodes (hubs). The SF network was generated with parameters $N=100$, $m_0=5$, and $m=3$ and the nodes' size represents their degree. {Additionally, the nodes are labeled in decreasing degree order. The parameters for the Rössler model are $a=0.01$, $b=0.2$, $c=30$, $D_\phi=D_\psi=0$, $D_\chi=0.2$, $\kappa=0.02$.}}
    \label{fig:Loc_L}
\end{figure*}
An initial attempt to understand the emergence of chimera states \cite{abrams_solvable_2008} involved analyzing a system composed of two populations of identical Kuramoto oscillators, which were strongly interconnected within each group and weakly connected between the two groups. Other {approaches} for identifying suitable network partitions into synchronized and desynchronized groups of oscillators have been developed by searching for specific network symmetries \cite{pecora_cluster_2014} or applying block diagonalization methods \cite{zhang_symmetry-independent_2020, zhang_mechanism_2021}. The objective of the techniques mentioned above is to analyze the network structure to identify parts of the network that will exhibit either coherence or incoherence. 
{Recently, the authors explored spectral properties as an indirect method to identify network substructures conducive to the emergence of chimeras \cite{symm_break}, through a symmetry method mechanism}. In this context, the authors have examined a model of coupled differential equations within a modular network. They demonstrated that by following a pattern formation mechanism, it is possible to accurately predict which module will exhibit synchronized or desynchronized oscillations. Although these symmetry-breaking methods \cite{Siebert, symm_break} explain the simultaneous existence of coherent and incoherent behavior, they are limited not only by the specificity of the structural features—strong modularity as in previous examples \cite{abrams_solvable_2008, pecora_cluster_2014, zhang_symmetry-independent_2020, zhang_mechanism_2021}—but also by the specificity of the model. 
Following a similar approach, in Ref. \cite{muolo_persistence_2024}, the authors leveraged the triangularity of the Laplacian eigenvectors matrix—common in most directed real networks \cite{Asllani2018PRE, asllani2018structure, Muolo2019, NN_stoch, baggio2020efficient, Johnson2020digraph, Muolo2021, OBrien_NN, nartallokaluarachchi2024broken}—to localize disorder within a subset of the network. {Albeit} such results represent an initial attempt to use pattern formation to explain chimera states, they lack generality in considering systems of coupled oscillators {where the modular structure is weaker or nonexistent.}
{Furthermore, pattern formation in networked systems has mostly focused on determining the conditions under which bifurcations occur, recent studies \cite{padmore2020modelling, pranesh2024effect, luo2024relationship} have explored pattern formation itself, with no reference so far to the emergence of chimeras.}
{In a different but related setting, Sethia et al. \cite{sethia_amplitude-mediated_2013,sethia_chimera_2014} proved the existence of amplitude-mediated chimeras in the nonlocal complex Ginzburg-Landau equation. These works highlight that different kinds of chimera states can be obtained, and probably not all of them are reducible to a single mechanism. However, they all share conditions related to the connectivity of the oscillators.}

Building on these insights, our work introduces an {alternative} perspective aimed at enriching the broader discourse on chimera states by applying pattern formation theory to the emergence of these states. {Such an approach is mathematically grounded on} weakly nonlinear analysis, {a technique used to study systems that are close to a linear regime but exhibit small nonlinearities,} originally developed by Kuramoto \cite{kuramoto_book}, and later applied {it also} to networked systems \cite{nakao_complex_2014, contemori_multiple-scale_2016, di_patti_ginzburg-landau_2018}. {This approach contrasts with the MSF formalism \cite{MSF}, which relies on the numerically calculated Lyapunov exponent and, as a result, does not generally permit direct analytical predictions of the final pattern to be observed} \footnote{Interestingly, in Ref. \cite{symm_break}, the authors studied a three-species model starting from a fixed point regime, similar to our approach in this work. However, there is a substantial difference between their paper and the current work. The former is focused on systems of coupled limit cycles with oscillations occurring at the node level, whereas in the latter, only traveling waves are possible. In the former, synchronized oscillations can occur in 'strong' modular networks, but no synchronization occurs in the general setting of complex networks considered in the current work.}. Specifically, it helps determine which parts of the network will desynchronize and which will remain synchronized. 
{The basics of weakly nonlinear analysis in this specific problem are as follows: first, identifying a small parameter that quantifies the strength of the nonlinearity in the system. This parameter should be small enough to ensure the system remains close to its linear state. The system begins in a homogeneous fixed-point regime near the Hopf bifurcation threshold, and a perturbation of the small parameter induces this bifurcation, triggering global oscillatory instabilities.} {Essentially, we could observe two distinct types of instabilities depending on the system: a complex non-spatial mode that influences node-level oscillations, and a spatial mode that, depending on its nature (real or complex), yields either stationary or oscillatory spatial patterns}. The crux of this approach, {that follows the same insights as Turing's instability analysis, and Turing's pattern formation,} is that the critical eigenfunction (or eigenvector) of the spatial mode is the determinant of the final form of the pattern at equilibrium. 

{Throughout this paper, for the sake of clarity and representativeness, we consider two models: the Brusselator, a set of autocatalytic reactions for two chemicals (two--species) that can exhibit complex dynamic behaviors, which we will show is related to the real spatial mode; and the Rössler attractor, a system of three nonlinear ordinary differential equations that serves as a paradigmatic example of a chaotic system. This is a three--species model related to the complex spatial mode}{. Details of both models are provided in the Appendix.}
 
{The key focus of this paper is to explore how the spectral properties of networks influence the development of chimera patterns, building upon the pattern formation theory that has been developed and matured over the past five decades \cite{cross_pattern_2009, Murray2008}.}
{Our insight comes from the fact that many random complex networks have strong eigenvector localization of the Laplacian matrix, i.e., the non-zero or dominant elements are concentrated in a small set of the components of the eigenvector \cite{mcgraw, nakao_loc, pastor2016distinct} (see Fig.~\ref{fig:Loc_L}(a), where the Laplacian eigenvectors are represented as columns in the visualization matrix for a Scale-Free (SF) network \cite{barabasi1999emergence}).}
 
This localization{, which} is similar to the eigenfunctions in {solid state physics, in the presence of impurities,} {and} known as Anderson localization \cite{anderson_absence_1958, bell1972dynamics, thouless1974electrons, kramer1993localization}{, has been observed to help understand disease spreading dynamics \cite{goltsev2012localization}}. In particular, it has been demonstrated that the localized entries correspond to a subset of nodes that share similar degrees \cite{nakao_loc}. Once one of these eigenvectors is selected as the critical one, {a supercritical bifurcation ensures that the emergent phase or/and amplitude {pattern} will follow the same trend observed in the localization structure}. Nodes with localized entries leave the synchronized manifold and change either their amplitude only, for stationary instabilities, or both the amplitude and the phase for oscillatory ones. In this paper, we will demonstrate that chimeras are structurally localized spatio-temporal patterns, as illustrated in Fig.~\ref{fig:Loc_L}(b). 

{The rest of the paper is structured as follows: in section II we briefly review the main ideas from {pattern} formation that we need to analyze the emergence of chimera states. In section III, we exemplify the use of the theory for two different dynamical systems in real and synthetic networks. Finally, we conclude discussing the implications of the performed analysis.}\\



{\section{Pattern formation in complex networks}}
\label{sec:PF}

{The theory of pattern formation investigates mechanisms that lead to the self-organization of complex patterns in reaction-diffusion systems. This theory provides insights into the principles underlying the spontaneous emergence of order in diverse biological and physical phenomena \cite{cross_pattern_2009, Murray2008}.
It is based on a weakly nonlinear analysis approach that allows predicting the shape of the final nonlinear patterns \cite{cross_pattern_2009, Murray2008, kuramoto_book}.} Originally developed to explore spatially extended patterns in continuous media \cite{cross_pattern_2009}, pattern formation theory has since been adapted to complex networks, significantly enhancing our insight into the dynamics of intricate patterns within these systems \cite{nakao_turing_2010, asllani_theory_2014, asllani_turing_2014, asllani_turing_2015, kouvaris_pattern_2015, petit_theory_2017, nakao_complex_2014, contemori_multiple-scale_2016}. In this section, we {briefly summarize the pattern formation theory} in a system of reaction-diffusion equations in networks of {coupled identical limit-cycle oscillators}, {following references} \cite{nakao_complex_2014, di_patti_ginzburg-landau_2018}. Indeed, Kuramoto used the pattern formation approach to derive his phase equation, approximating the coupling of oscillators with a continuous Laplacian operator. In contrast, our approach incorporates full amplitude-phase models, thus broadening the applicability of pattern formation theory to encompass both phase and amplitude chimeras. Like Kuramoto {did}, we demonstrate that changes in the amplitude of the pattern near criticality are indeed negligible.

{Let us} start by considering a {\(M\times N-\)}dimensional reaction-diffusion system and label \({\bf x}_j(t)\) for \(j=1, \ldots, N\) as the {M}-dimensional vector representing densities of the chemical or species under consideration, where the index \(j\) refers to the node of the network. Globally, the dynamics of the system {is} described by the set of networked coupled differential equations.
\begin{equation}\label{eq:systemRD} 
\dot{\bf x}_j = {\bf F} ({\bf x}_j , \boldsymbol{\mu}) +{\bf  D} \sum_{k=1}^N {L_{jk}}{\bf x}_k,\;\;\; \forall j
\end{equation}
where the {M}-dimensional vector function \({\bf F}\) specifies the nonlinear reaction part with \(\boldsymbol{\mu}\) representing the vector of parameters. For the spatial part, diffusive coupling: \({\bf D}= \kappa\, {\text{diag}(D_1,D_2, \dots, D_M)}\) denotes the diagonal matrix of diffusion coefficients where \(\kappa\) is a constant parameter representing the coupling strength. {The network structure is encapsulated in {${\bf L}$},} the Laplacian kernel defined through the adjacency matrix ${\bf A}$ as ${\bf {L}} = {\bf A} - {\bf K}$, where ${\bf K}$ is the diagonal matrix of the degrees. The system \eqref{eq:systemRD} admits a homogeneous equilibrium point ${\bf x}^*$. It is also required that when {only the reaction part is considered} (\(\kappa=0\)), \({\bf x}^*\) undergoes a Hopf bifurcation for \(\boldsymbol{\mu} = \boldsymbol{\mu}_0\). Slightly above this threshold, \({\bf x}^*\) becomes unstable, and a limit cycle emerges for each node of the network in its {non-spatial} limit (\(\kappa=0\)). The spatial coupling (\(\kappa\neq 0\)) is known to synchronize these limit cycles in a globally coherent synchronization across the entire network. However, if global instability conditions are fulfilled, tiny non-homogeneous perturbations can destabilize the uniform synchronous equilibrium.

{We {begin} by introducing small inhomogeneous perturbations, {\({\bf u}_j\)}, to the uniform equilibrium point, {thereby slightly perturbing the system from its steady state, resulting in} {\({\bf x}_j={\bf x}^*_j+{\bf u}_j\)} for \(j=1, \ldots, N\). Substituting this into equations \eqref{eq:systemRD} and performing a linearization yields the following equation for the short time evolution of \({\bf u}_j\):
\begin{equation}\label{eq:lin_eq}
\dot{{ \bf u}}_j \approx {\bf  {J}  u}_j  +  {\bf D} \sum_{k=1}^N {L_{jk}} {\bf u}_k,
\end{equation}
where {\({\bf J}\)} is the Jacobian matrix evaluated at the steady state \({\bf x}^*\). 
Given that the Laplacian eigenvectors form an orthogonal basis, we look for solutions of the form:
\begin{equation}
\mathbf{u}_j = \sum_{\alpha=1}^N \mathbf{c}_\alpha {V_{j}^{\alpha}} e^{\lambda_\alpha t},
\label{eq:soln}
\end{equation}
where \(\mathbf{c}_\alpha\) are \(M\)-dimensional vectors representing the coefficients for each species, {\(V_{j}^{\alpha}\)} are the entries of the Laplacian eigenvectors, and \(\lambda_\alpha\) are the growth rates of the perturbations, all corresponding to the eigenmode {\(\Lambda^\alpha\)}. Substituting this solution into Eq. \eqref{eq:lin_eq} and after some algebraic manipulation (see Appendix for details), using the eigenvalue/eigenvector property \(\sum_{k=1}^N {L_{jk}} {V_{k}^{\alpha}} = {\Lambda^{\alpha}} {V_{j}^{\alpha}}\), one can establish that the growth rates $\lambda_\alpha$ are the $M$ eigenvalues of \(({\mathbf{J}} + {\Lambda^{\alpha}} \mathbf{D})\). The relationship \(\lambda_\alpha({\Lambda^{\alpha}})\), mapping the Laplacian eigenvalues $\Lambda^{\alpha}$ to the growth rate $\lambda_\alpha$ with the maximum real part among the $M$ eigenvalues, is also known as the discrete dispersion relation \cite{nakao_turing_2010, asllani_theory_2014}, analogous yet distinct from the MSF.} {Before proceeding further, we want to point out that, although the assumption of identical oscillators is a requirement for the dispersion relation/MSF formalism and is commonly used in the literature, it has been shown that such a framework can be extended to relax this constraint \cite{Sun_2009, nazerian2023synchronization}.} 

\begin{figure*}
    \centering
    \includegraphics[width=\textwidth]{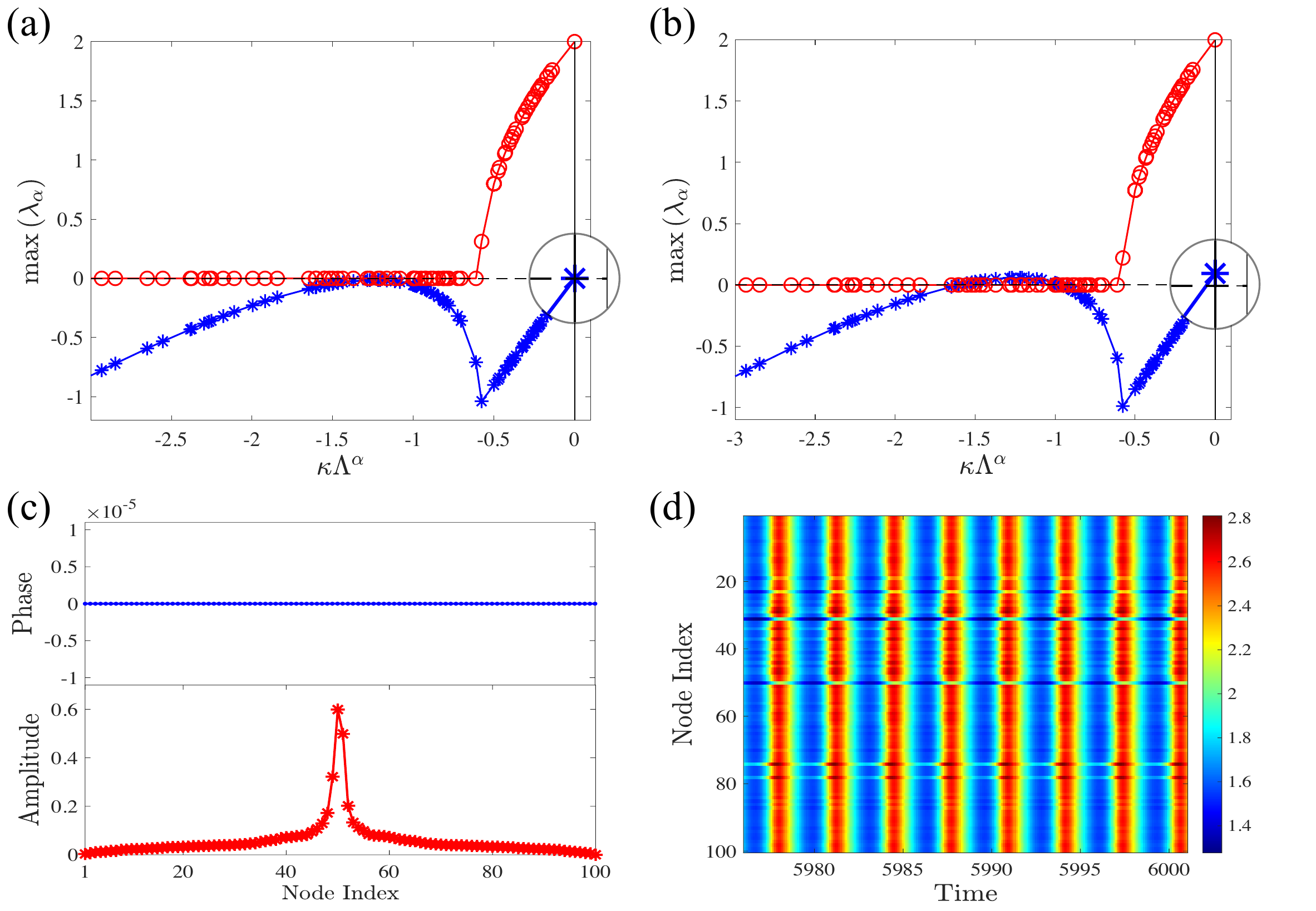}
    \caption{{\textbf{Amplitude Chimera States in the Brusselator Model.} Real (blue stars) and imaginary (red circles) parts of the dispersion relation: (a) with \(b=5\) in the pre-perturbation setting and (b) with \(b=5.1\) where chimera occurs, zoomed at the origin for clarity. (c) Phase (upper) and amplitude (lower) differences from the original limit cycle, with oscillators maintaining the same phase. To better visualize the localization of the disorder, the nodes are intentionally indexed to yield a {center-peaked symmetric} amplitude distribution. (d) Pattern evolution at equilibrium, showing amplitude disruptions. {The remaining parameters} are $a=2$, $D_\phi=5$, $D_\psi=13$, and $\kappa=0.1$, and the network used is depicted in Fig. \ref{fig:Loc_L}.}}
    \label{fig:Bruss}
\end{figure*} 

\vspace{.25cm}

{As mentioned already, Eq. \eqref{eq:soln} describes the behavior of the system under scrutiny only for the initial evolution of the system near the (unstable) steady state, which in principle differs from the long-term behavior, i.e., equilibrium, where the final pattern is observed. An exception to this occurs when the parameters are set as described earlier, near the critical point and for weak spatial coupling, i.e., \(\kappa \ll 1\). This regime allows for a weakly nonlinear analysis through multiple-scale perturbation theory, providing an equation to {approximately} describe the amplitude evolution of the perturbations, known as the Complex Ginzburg-Landau (CGL) equation. The CGL equation serves as a normal form for describing the amplitude of a pattern} \footnote{{It is worth noting that numerical studies have shown that the CGL equation can exhibit chimera states \cite{sethia_amplitude-mediated_2013, sethia_chimera_2014}. This provides further evidence that, as a first-order approximation of the general reaction-diffusion system \eqref{eq:systemRD}, chimera states can theoretically appear in any system of coupled oscillators given an appropriate choice of parameters.}}. 
{In this paper, we will not delve into the details of analytical/numerical approximations of equation \eqref{eq:systemRD}, but instead, we will use the CGL formalism to validate the pattern predictions obtained through linear stability analysis. Details of the derivation of the CGL amplitude equation are given in the Appendix. Notably, the significance of the weakly nonlinear approach lies in its role as the normal form for a Hopf pitchfork bifurcation, particularly in the case of a supercritical bifurcation. For a suitable choice of parameters, the linear stability analysis of the CGL solution indicates stability in the supercritical case and instability in the subcritical case. This aligns with the specific scenario where, as predicted by perturbation theory, a small amplitude pattern is expected as operating slightly above the critical point \(\boldsymbol{\mu_0}\). In the literature, it has been established that the CGL is always in the supercritical regime for the Brusselator model \cite{di_patti_ginzburg-landau_2018}. Additionally, we numerically demonstrate in the Appendix that this is consistently also true for the Rössler model with our selected parameters.}

This point is crucial and conclusive for the pattern analysis. The normal form, described by the Complex Ginzburg-Landau equation, reveals that near the threshold of instability, the solution of the original reaction-diffusion system \eqref{eq:systemRD} reaches equilibrium after a rapid saturation of the linear evolution of the perturbation, $\mathbf{u}_j$, confirming Turing's original intuition \cite{turing_chemical_1990}. Consequently, the final pattern will take the form of the critical eigenvector (or a linear combination of such eigenvectors) established {using} linear stability analysis of \eqref{eq:systemRD} as follows:
\begin{equation}
\mathbf{u}_j = \sum_{\alpha \in \mathcal{U}} \mathbf{c}_\alpha {V_{j}^{\alpha}} e^{\lambda_\alpha t} + \sum_{\alpha \in \mathcal{S}} \mathbf{c}_\alpha {V_{j}^{\alpha}} e^{\lambda_\alpha t},
\label{eq:lin_exp}
\end{equation}
{where we have separated Eq.(\ref{eq:soln}) in $\mathcal{U}$, the set of indices corresponding to unstable modes, and $\mathcal{S}$ the set of indices corresponding to stable modes. Note that}, $\alpha = N$ corresponding to the null Laplacian eigenvalue ${\Lambda^N}=0$ (the {non-spatial} mode) belongs to $\mathcal{U}$ since $\lambda_N = i \omega_0$.

Two scenarios arise from this analysis: (a) \textit{stationary spatial instability}, when the critical spatial mode derived from the linear stability analysis has no imaginary part, $\Im{(\lambda_\alpha)} = 0 \ \forall \alpha \in \mathcal{U} \setminus \{N\}$. In this case, the spatial mode contributes only to the shape of the final pattern, not {to} its temporal dynamics; and (b) \textit{oscillatory instability}, when the imaginary part of at least one of the critical spatial modes is non-zero, $\exists \alpha = c \in \mathcal{U} \setminus \{N\}$ such that $\Im{(\lambda_\alpha)} \neq 0$. In this case, {an additional} temporal frequency $\omega_c = \Im{(\lambda_\alpha)}$ (compared to the intrinsic oscillations $\omega_0$) is added to the dynamics of the pattern at equilibrium.\\


{\section{Amplitude and phase localization}}
\label{sec:Local}

Building on the spectral localization concepts discussed earlier in this paper, we now examine the implications of the outcomes derived from {the linear stability} analysis. This analysis suggests that depending on the nature of the dispersion relation {introduced earlier}, the system may exhibit {stationary spatial instabilities, and oscillatory instabilities. Our claim is that these solutions do correspond to }various types of chimera states, characterized by either amplitude or phase disorder, or both. Notably, the eigenvectors ${\bf {V}^{\alpha}},\,\alpha \in \mathcal{U}$ are critical to our study, {as we will show on specific examples.} 

{We begin by considering the dispersion relation for two well-know dynamical systems in networks: the Brusselator {model, and the Rössler model (see Appendix for the models' details). The Brusselator model dispersion relation is }illustrated in Figure~\ref{fig:Bruss}}. {They represent respectively the pre-perturbation settings Fig.~\ref{fig:Bruss}(a) and the post-perturbation settings Fig.~\ref{fig:Bruss}(b).} 
It can be noticed that the Brusselator-based reaction-diffusion system exhibits a Hopf bifurcation only at the origin, identified by the presence of a non-null imaginary part (red circles), which leads to the intrinsic oscillations of the nodes. 

Mathematically, this is explained by the fact that being a two-species model, the only kind of instability it can undergo at the spatial level, i.e., \(\Lambda^{\alpha} \neq 0\), is of a stationary type where the set of eigenvalues for which the instability occurs are all strictly real \cite{cross_pattern_2009, Murray2008}. Our analysis predicts no traveling wave effects in this model, as the critical non-zero mode of the Laplacian is purely real. Therefore, oscillations are confined to the {non-spatial} {mode} at the origin. Consequently, the pattern governed by the localized (purely real) entries of the critical eigenvectors \({\bf{V}^{\alpha}}\) will display amplitude disorder at a subset of nodes while maintaining their original phase in both frequency and phase lag. {This is  confirmed in Fig.~\ref{fig:Bruss}(c),} where the amplitude-only chimera state is distinctly visible. The indexing of nodes is adjusted to represent a {center-peaked symmetric} distribution. {Such representation aims to better visualize the localization of amplitude and/or phase disorder by separating the synchronized from the unsynchronized oscillators on one side and offering a better comparison of the amount of disorder per node.} Fig.~\ref{fig:Bruss}(d) shows the temporal evolution of the amplitude chimera at equilibrium. 

\begin{figure*}
    \centering
    \
    \includegraphics[width=\textwidth]{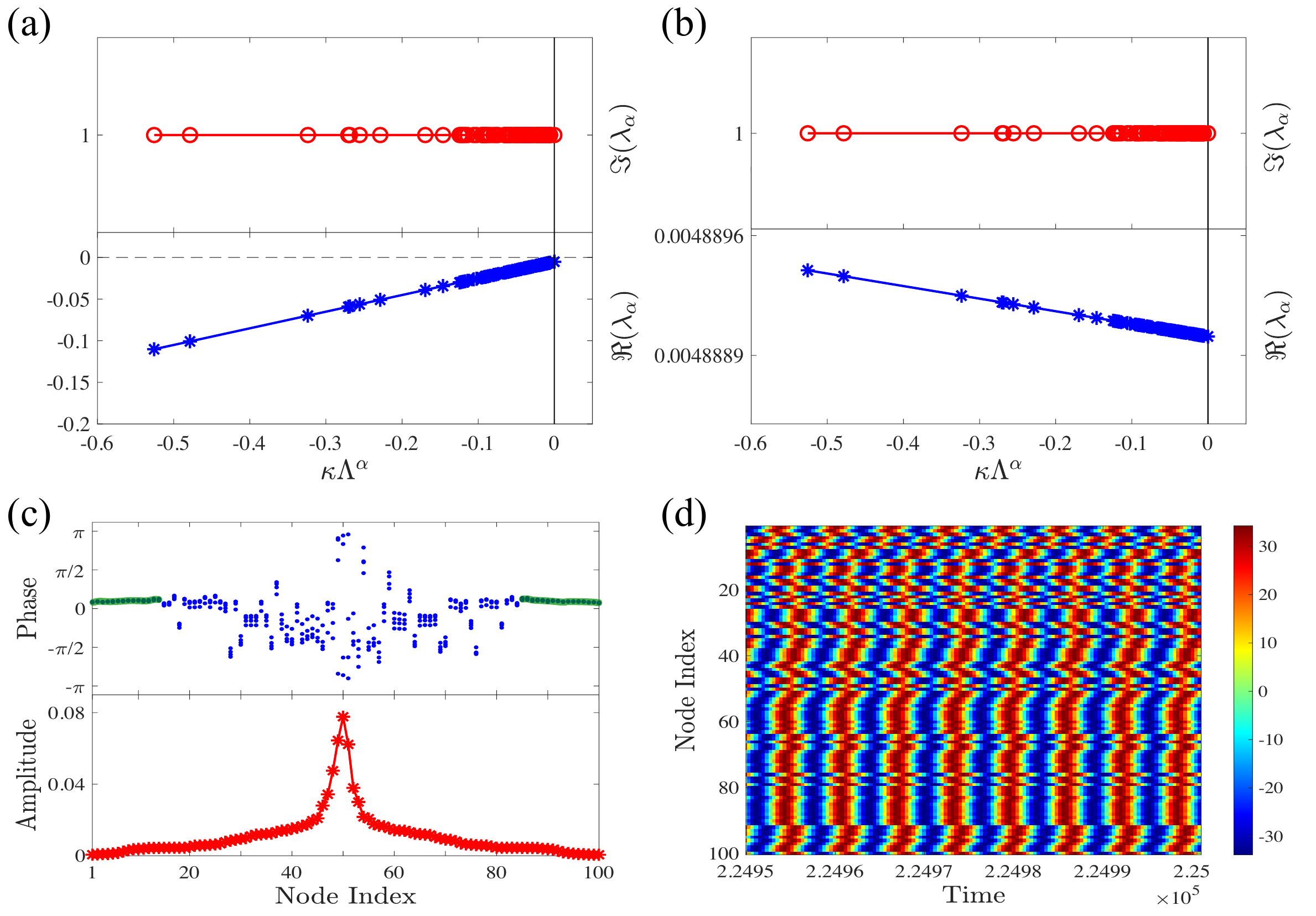}
    \caption{{\textbf{Chaotic Amplitude-Phase Chimera States in the Rössler Model.} Real (blue stars) and imaginary (red circles) parts of the dispersion relation: (a) with \(a=-0.01\) and \(D_\phi=D_\psi=0.2\) in the pre-perturbation setting and (b) with \(a=0.01\) and \(D_\phi=D_\psi=0\) where chimera occurs, zoomed at the origin for clarity. (c) Stroboscopic analysis of phase differences (upper) reveals chaotic, unsynchronized oscillators, with phase-synchronized ones highlighted in green. Amplitude differences (lower) show a localized pattern. In both cases, relabeling the nodes with a {center-peaked symmetric} distribution helps visualize the localization of amplitude and phase disorder. (d) Equilibrium pattern showing phase disruptions. {The remaining parameters are} $b=0.2$, $c=30$, $D_\chi=0.2$, $\kappa=0.02$, and the network used is depicted in Fig. \ref{fig:Loc_L}.}}
    \label{fig:Rosschaos}
\end{figure*}

\vspace{.25cm}

\begin{figure*}
    \centering
    \
    \includegraphics[width=\textwidth]{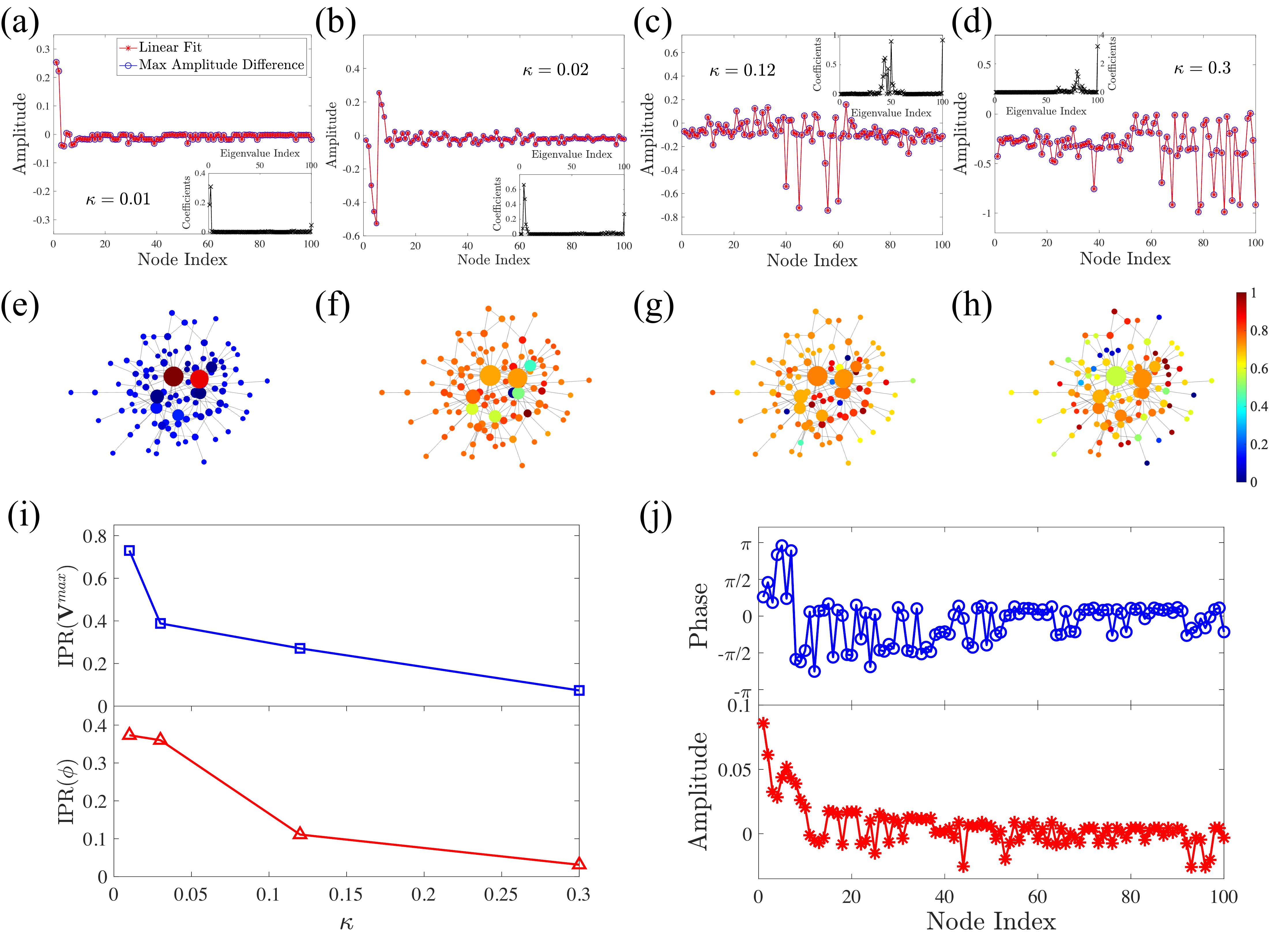}
    \caption{{\textbf{Pattern localization.} (\textit{Brusselator}) (a)--(d) Snapshots of the final amplitude pattern with blue circles, for different values of the coupling strength $\kappa$ ({also} specified in panel (i)). In the insets, coefficients from the linear fitting of eigenvectors, depicted as black crosses, are confirmed by the pattern's reconstruction with red stars in the main figure. (e)--(h) Network representation of the patterns above confirms the localization in nodes with similar degrees (here represented with similar size). (i) IPR calculation for the most critical Laplacian eigenvector (upper) vs. the IPR of the final patterns (lower). (\textit{Rössler}) (j) The upper section displays phase differences from the original limit cycle (blue circles), while the lower part shows amplitude differences (red stars). The setting of parameters is as described in Fig. \ref{fig:Bruss} and Fig. \ref{fig:Rosschaos}, respectively, with the only difference that \(D_\psi\) for the Brusselator is adjusted in the interval $[12.99, 13.2]$.}} 
    \label{fig:Bruss_fit}
\end{figure*}

\vspace*{.25cm}

The situation differs drastically in the Rössler model (a three-species model), where a Hopf bifurcation can occur for a spatial model different from \({\Lambda^{\alpha}} \neq 0\) \cite{cross_pattern_2009, Murray2008}. 
As anticipated earlier, such instability occurs when the non-null spatial mode of the Laplacian has a corresponding Jacobian eigenvalue in the dispersion relation with an imaginary part different from zero. {This is shown in the dispersion relation of Fig. \ref{fig:Rosschaos}, where a small change in the parameters transitions the system from the state presented in Fig.~\ref{fig:Rosschaos}(a) to the one in Fig.~\ref{fig:Rosschaos}(b), where the modes with a positive real part (blue stars) now have a nonzero imaginary part (red circles).} This mechanism, known as oscillatory instability \cite{cross_pattern_2009}, is responsible for generating traveling waves in the spatial support. The peculiar characteristics of the Rössler model, known for its oscillatory behavior—whether chaotic or not—result in all modes of the dispersion relation transitioning from stable to unstable, with a decreasing slope towards the origin as shown in Fig.~\ref{fig:Rosschaos}(b). This, together with an omnipresent imaginary part in the dispersion relation, suggests that a spatially varying oscillatory pattern can be anticipated. In the parameter setting shown in Fig.~\ref{fig:Rosschaos}(c), we observe how initially synchronized but chaotic states emerge, exhibiting localized amplitude disorder. Due to its small amplitude, this disorder can be considered negligible when near the instability threshold, as anticipated by Kuramoto \cite{kuramoto_book}. This is understandable considering the small amplitude approach to the problem on one hand {(see Appendix)}, and the two-speed pace of the imaginary and real parts of the dispersion relation which allows a full development of the oscillations while the amplitude develops slightly, so one can focus on the phase only. Notably, the phase behavior, as depicted in a stroboscopic plot, {Fig.~\ref{fig:Rosschaos}(c),} shows localized chaotic incoherence within a subgroup of oscillators, maintaining an average phase difference from the originally synchronized state. As before, the indexing of nodes is adjusted to represent a {center-peaked symmetric} distribution for both phase and amplitude differences {for a better visualization of the separation of the coherent and incoherent oscillators. The same behavior is confirmed by the temporal evolution in Fig.~\ref{fig:Rosschaos}(d).}

{Complementary comparisons between the set of critical eigenvectors and the final pattern confirm the precise localization of the amplitude of the pattern. Figure \ref{fig:Bruss_fit}(a)-(d), display snapshots of the Brusselator equilibrium pattern shown with blue circles, for different values of the coupling strength \(\kappa\). These patterns are analytically reconstructed using a linear combination of the basis of the Laplacian eigenvectors, expressed as $\boldsymbol{\phi}=\sum_{\alpha=1}^N m_\alpha {\bf{V}^{\alpha}}$, where the scalar coefficients $m_\alpha$ have been numerically estimated. As observed in the {insets, the coefficients peak at specific values (corresponding to the critical eigenvalues of their respective dispersion relations), confirming} that only a few eigenvectors shape the final patterns. Additionally, there is a second peak corresponding to the origin, where the eigenvector is uniformly distributed.
Importantly, the indexing of the nodes in Figure \ref{fig:Bruss_fit} remains unchanged compared to Fig.~\ref{fig:Loc_L}(a), meaning that nodes are labeled in decreasing order of their degrees. Following the portrayal of Fig.~\ref{fig:Loc_L}(a), the localization is stronger in the first eigenvectors of the Laplacian and becomes weaker as their order increases. We can control the critical eigenvector by tuning the coupling strength \(\kappa\). As anticipated previously, the eigenvector localization is centered around entries that correspond to nodes with similar degrees, meaning that the amplitude disorder will localize accordingly, as shown in Fig.~\ref{fig:Bruss_fit}(e)-(h). To quantify the amount of localization, in Fig.~\ref{fig:Bruss_fit}(i) we have used the Inverse Participation Ratio (IPR), initially developed to measure the Anderson localization, defined for a non-normalized vector \(\boldsymbol{\Phi}\) with components \(\Phi_i\) as \( \text{IPR}(\boldsymbol{\Phi}) = {\sum_{i} \Phi_i^4}/{\left(\sum_{i} \Phi_i^2\right)^2} \) \cite{anderson_absence_1958}. The upper part shows the IPR of the most critical eigenvector and the lower part shows the IPR of the amplitude pattern for different values of \(\kappa\) used in the previous panels. The results confirm the increasing disorder as \(\kappa\) increases.}

{Similarly, it is shown in Fig.~\ref{fig:Bruss_fit}(j) that for the Rössler model the localization decreases with the node index, following the initial node indexing of Fig.~\ref{fig:Rosschaos}(a). We avoid the linear fitting of eigenvectors because, as indicated by the dispersion relation, all the Laplacian modes are critical. Nevertheless, the figure reveals a clear trend of amplitude-phase differences, exhibiting the same decrement for increasing indices as observed in the corresponding real part of the dispersion relation. This suggests that nodes with higher real parts of the critical eigenvalues, $\lambda_\alpha$, are more significantly disturbed than those with lower values. Considering that the nodes are ordered according to decreasing degree, this explains why in Fig.~\ref{fig:Loc_L}(b) the disorder is localized in the hubs of the network, a phenomenon already observed in other settings \cite{arenas2006synchronization}.}\\

\begin{figure*}
    \centering
    \
    \includegraphics[width=\textwidth]{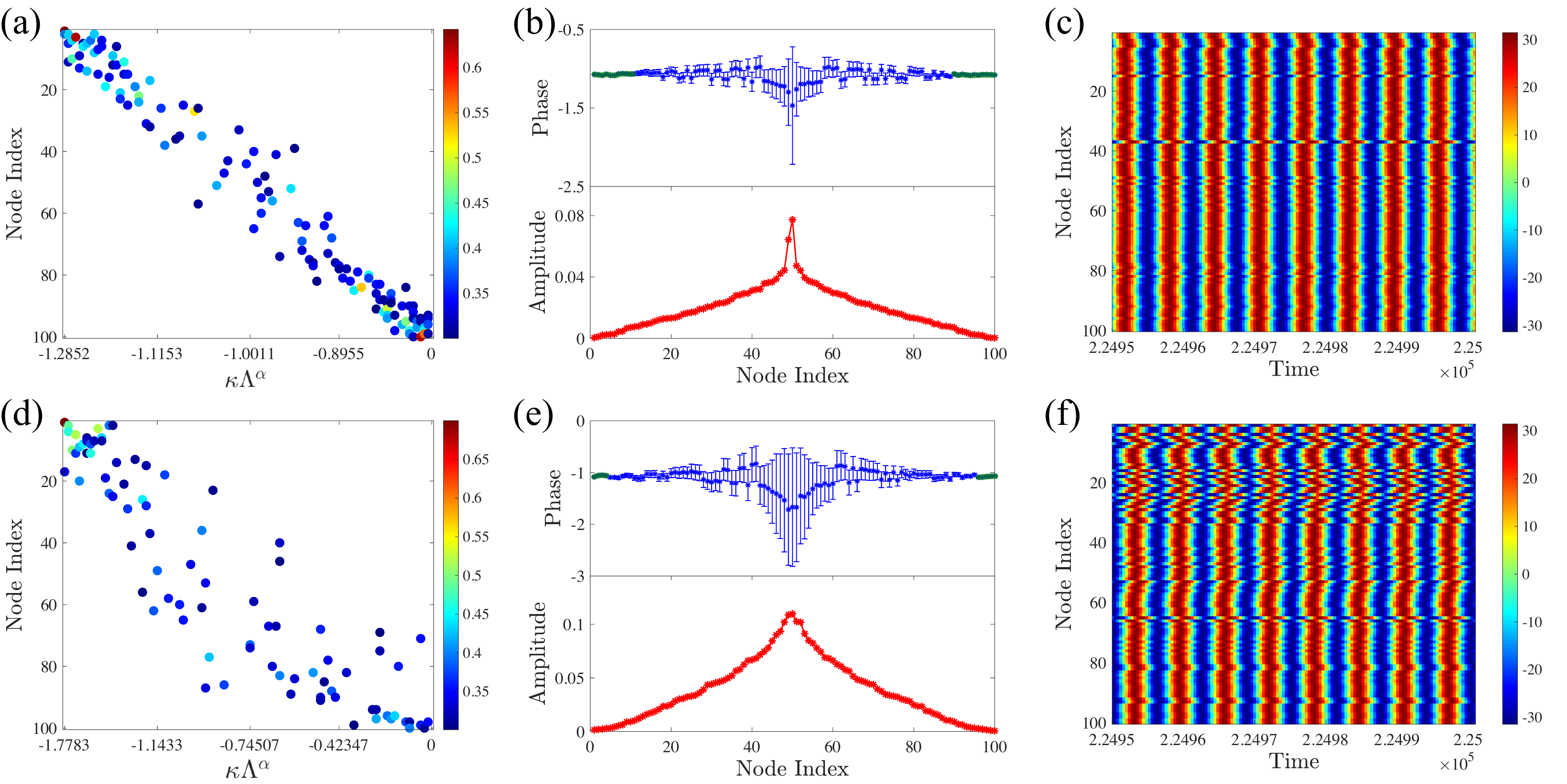}
    \caption{\textbf{Chaotic Amplitude-Phase Chimera States in the R\"{o}ssler Model for the Erdös-Rényi (upper) and Small-World (lower) networks.} {Panels (a) and (d) represent the eigenvector localization, (b) and (e) show the amplitude-phase chimeras, and (c) and (f) depict the pattern evolution at equilibrium for the species \(\phi\), respectively.} Common parameters for both cases are $a=0.002$, $b=0.2$, $c=30$, $D_\phi=D_\psi=0$, $\kappa=0.004$. For the Erd\H{o}s-Rényi network, $D_\chi=0.5$, and we have generated a 100-node network with a wiring probability $p=0.5$. For the Small-World network, $D_\chi=5$, and the network has 100 nodes, generated starting from a ring of $k=4$ near neighbors and a rewiring probability $p=0.5$.}
    \label{fig:Ross_ER_SW}
\end{figure*}

\begin{figure*}
    \centering
    \
    \includegraphics[width=\textwidth]{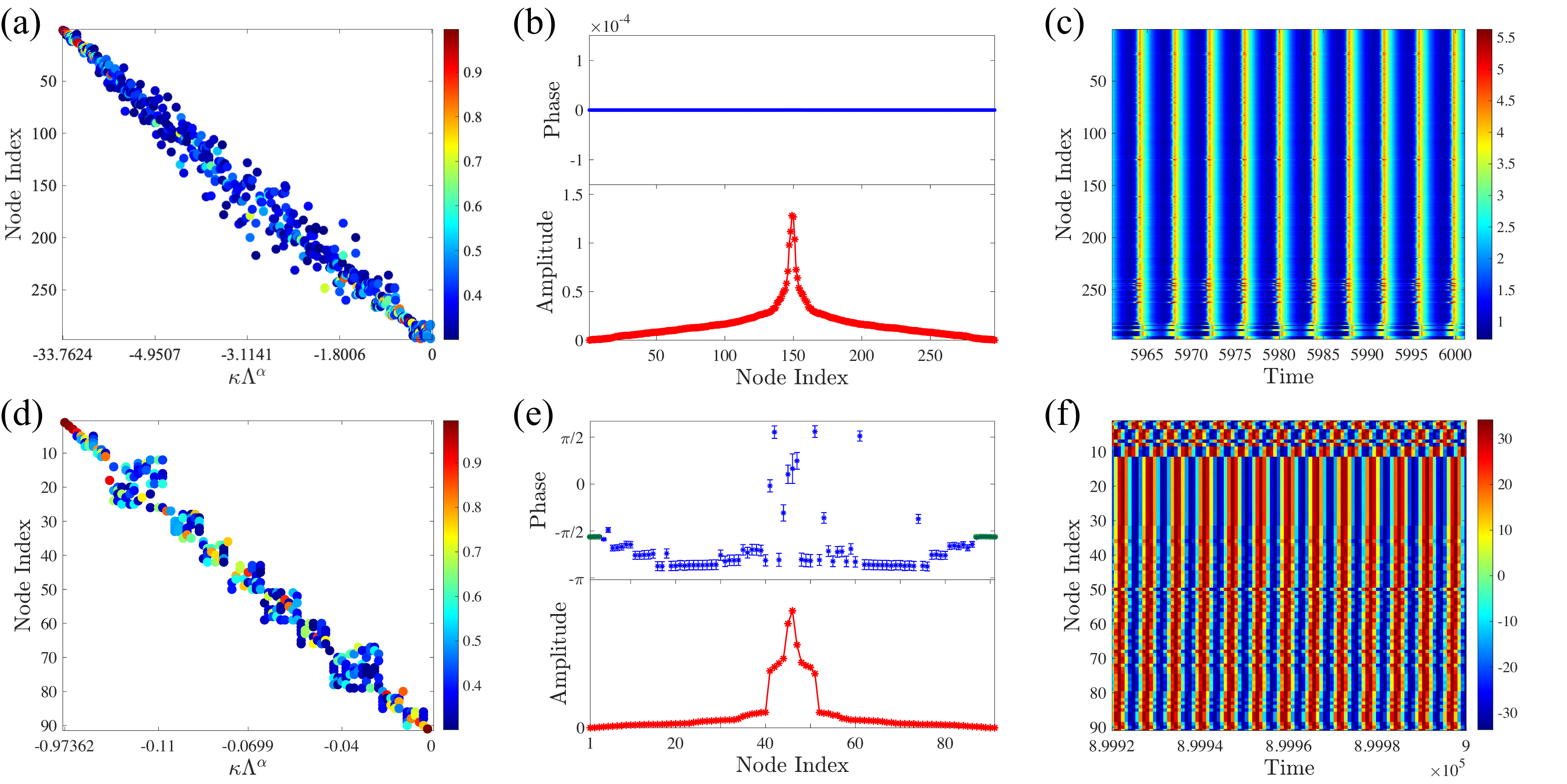}
    \caption{\textbf{Chaotic Amplitude-Phase Chimera States in the Brusselator for the \textit{C. elegans} (upper) and the Rössler \textit{Rhesus macaque} (lower) neural networks.} {Panels (a) and (d) represent the eigenvector localization, (b) and (e) show the amplitude-phase chimeras, and (c) and (f) depict the pattern evolution at equilibrium for the species \(\phi\), respectively.} For the Brusselator model, the parameters are $a=2$, $b=5.5$, $D_\phi=1$, $D_\psi=2.56$, and $\kappa=0.5$. For the Rössler model, the parameters are $a=0.01$, $b=0.2$, $c=30$, $D_\phi=D_\psi=0$, $D_\chi=0.025$, and $\kappa=0.02$.}
    \label{fig:Ross_real}
\end{figure*}

{\subsection{Amplitude-phase chimera states in other synthetic and real-world networks}}
\label{subsec:Other}

{In this part, we extend and validate our results by providing further evidence that localized (chaotic) disordered oscillatory patterns are also possible in other network topologies, such as Erd\H{o}s-R\'enyi (ER) and Small-World (SW) networks as demonstrated in Figure \ref{fig:Ross_ER_SW}. The randomness of the connections in these structures leads to the localization of the eigenvectors within a subset of nodes, similar to what has been presented so far. As a consequence, amplitude and phase chimera patterns are also observed in these topologies. Our approach is also validated for empirical networks, Figure \ref{fig:Ross_real},  particularly neuronal networks, where both empirical observations of chimeras and the role of synchronization as a main functioning mechanism have been documented. By applying our method to these real-world networks, we aim to corroborate the results obtained throughout this paper, showing that the observed patterns are not only theoretically feasible but also practically significant. This validation highlights the broader applicability and robustness of our findings across different network types and underscores their relevance in understanding the complex dynamics of real-world systems.}

{Unlike the previous figures, which use the stroboscopic representation of the chaotic behavior of the chimera states, here we have chosen to represent the maximum difference in the phases from their mean value using error bar notation. In Fig.~\ref{fig:Ross_ER_SW}(a)-(c), we have shown the amplitude-phase localization for an {ER} network. It can be noticed that the amplitude, although of a magnitude negligible compared to the phase, has weaker localization, whereas the phase chimera shows similar behavior as for the scale-free case in the previous section. In Fig.~\ref{fig:Ross_ER_SW}(d)-(f), the simulation for the SW network is presented. Here, not only do the amplitude perturbations show less localization, but the phase ones also show less localization compared to the ER case. The interpretation for such a quantitative difference from the SF network can be found in the localization properties of the eigenvector, recalling that for the particular case of the Rössler model, all the modes are critical with decreasing monotonicity toward the origin respective growth rate. As before, it is more likely that the eigenvectors with lower Laplacian eigenvalues manifest more, and in fact, for the SF network, such eigenvectors are more localized in this range of the spectrum compared to the previous two scenarios we saw here.}\\
\indent
{For the two empirical connectomes considered in Fig. \ref{fig:Ross_real}, namely \textit{C. elegans} (using the Brusselator model) and macaque \textit{Rhesus brain} (using the Rössler model), the localization is overall higher across the range of the Laplacian eigenvalues, which is also reflected in the amplitude chimera localization. Notice that for the macaque connectome, the chaoticity manifests weaker than previously, and disorder in the phase has a high range but low average magnitude, a result expectable considering the marked localization of the Laplacian eigenvectors.}\\


{\section{Conclusions and discussion}}
\label{sec:Conlus}

In this paper, we explored the emergence of chimera states in complex networks through the lens of pattern formation theory. {On the one hand}, chimera states, far from being merely mathematically elegant, play a crucial role in understanding synchronization patterns across diverse systems, from the brain to ecological networks \cite{chimera_brain, chimera_firefly}. On the other hand, pattern formation theory has seen significant development in recent decades, offering insights into the formation of complex spatial patterns in both fluid and solid systems, with applications spanning physics \cite{cross_pattern_2009} and biology \cite{Murray2008}. Pattern formation theory, through {stability analysis building on nonlinear perturbation theory}, enables the prediction of patterns in nonlinear equilibria.
Since 2010, the study of Turing patterns in complex networks \cite{nakao_turing_2010} has sparked numerous investigations into pattern formation in various network structures. 
Despite the advancement in pattern formation theory in networks, applications have often been constrained to specific network types 
such as modular \cite{symm_break} or non-normal ones \cite{muolo_persistence_2024}, which exhibit spectral localization due to clustering or triangularization of Laplacian eigenvectors entries. 

{However, a general analysis of chimera state formation in complex networks has been lacking. In this manuscript, we aim to fill this gap. By leveraging the fact that many complex networks possess strongly localized eigenvectors of the Laplacian, we propose a slight destabilization of the coupled oscillator system. By tracking the propagation of this destabilization through the localized entries of the critical Laplacian eigenvector, we determine how disorder in amplitude and/or phase is induced, while synchrony is maintained in the rest of the network.}\\ 
{This approach} provides a systematic explanation for the emergence of chimera patterns from the initial time to equilibrium, aiming in this way to fill a gap in the existing literature. While the authors acknowledge that further efforts are needed to connect the results of this paper with phase-reduced models, such as the Kuramoto model, where chimeras were originally observed and subsequently studied, they are optimistic that this {proposed} approach will pave the way for interesting future findings. These include exploring the multiple different chimera patterns known to exist in the literature \cite{zakharova_chimera_2020}, elucidating the peculiar conditions under which such states arise, and last but not least, offering tools for analyzing their stability.\\


{\section*{Appendix}}


{\subsection{Methods}}
\subsubsection{{Details of the} Linear Stability Analysis (LSA)}
\label{eq:LSA}

The linearized reaction-diffusion system is:
\[
\dot{\mathbf{u}}_j = {\mathbf{J}} \mathbf{u}_j + \mathbf{D} \sum_{k=1}^N {L_{jk}} \mathbf{u}_k, \quad \forall j,
\]
where {$\mathbf{u}_j$ is the M-dimensional vector of the perturbation for the node $j$.} We expand $\mathbf{u}_j$ in terms of {the basis of eigenvectors} of the Laplacian:
\[
\mathbf{u}_j = \sum_\alpha \mathbf{c}_\alpha {V_{j}^{\alpha}} e^{\lambda_\alpha t},
\]
where $\mathbf{c}_\alpha$ are {M}-dimensional vectors representing the coefficients for each species, {$V_{j}^{\alpha}$} are the entries of the eigenvectors, and $\lambda_\alpha$ are the growth rates of the perturbations.

Substituting this expansion into the linearized reaction-diffusion equation:
\begin{align*}
\sum_\alpha \lambda_\alpha \mathbf{c}_\alpha {V_{j}^{\alpha}} e^{\lambda_\alpha t} = {\mathbf{J}} \sum_\alpha &\mathbf{c}_\alpha {V_{j}^{\alpha}} e^{\lambda_\alpha t}+\\ &+ \mathbf{D} \sum_{k=1}^N {L_{jk}} \sum_\alpha \mathbf{c}_\alpha {V_{k}^{\alpha}} e^{\lambda_\alpha t}.
\end{align*}
Using the property of the eigenvalues and eigenvectors of ${\bf {L}}$ we have
\(
\sum_{k=1}^N {L_{jk}} {V_{k}^{\alpha}} = {\Lambda^{\alpha}} {V_{j}^{\alpha}},
\)
so we get:
\begin{align*}
\sum_\alpha \lambda_\alpha \mathbf{c}_\alpha {V_{j}^{\alpha}} e^{\lambda_\alpha t} = {\mathbf{J}} \sum_\alpha &\mathbf{c}_\alpha {V_{j}^{\alpha}} e^{\lambda_\alpha t}+\\ & + \mathbf{D} \sum_\alpha \Lambda^{\alpha} \mathbf{c}_\alpha {V_{j}^{\alpha}} e^{\lambda_\alpha t}.
\end{align*}
Equating the terms with the same exponential factor $e^{\lambda_\alpha t}$:
\[
\lambda_\alpha \mathbf{c}_\alpha {V_{j}^{\alpha}} = {\mathbf{J}} \mathbf{c}_\alpha {V_{j}^{\alpha}} + \mathbf{D} {\Lambda^{\alpha}} \mathbf{c}_\alpha {V_{j}^{\alpha}}.
\]
Since this relationship must hold for each node $j$, and assuming that for at least one node $j$, ${V_{j}^{\alpha}} \neq 0$, we get:
\[
\lambda_\alpha \mathbf{c}_\alpha = ({\mathbf{J}}+ {\Lambda^{\alpha}} \mathbf{D}) \mathbf{c}_\alpha.
\]
The growth rates $\lambda_\alpha$ are the eigenvalues of the matrix $({\mathbf{J}} + {\Lambda^{\alpha}} \mathbf{D})$. The dispersion relation for the growth rates $\lambda_\alpha$ as a function of the eigenvalues of the Laplacian matrix $\lambda_\alpha\left({\Lambda^{\alpha}}\right)$.\\


\subsubsection{Definitions of the Brusselator and Rössler Models}
\label{sec:models}

\textit{\underline{Brusselator Model}}\\
The Brusselator is a two-variable model of an autocatalytic reaction, and can be considered as a single, uncoupled node from a network perspective. It is described by the following differential equations:
\[
\begin{aligned}
\frac{dx}{dt} &= a + x^2 y - (b + 1) x, \\
\frac{dy}{dt} &= b x - x^2 y,
\end{aligned}
\]
where $x$ and $y$ represent the concentrations of two chemical species, and $a$ and $b$ are constants. The fixed points of the system are given by \( (x^*, y^*) = (a, b/a) \). This model exhibits a Hopf bifurcation, transitioning from a stable fixed point to a limit cycle when the parameter $b$ exceeds the critical threshold $b_c = 1 + a^2$, indicating the onset of oscillatory behavior.

\textit{\underline{Rössler Model}}\\
The Rössler model is a three-variable system that can also be seen as a single, uncoupled node in a network context, primarily used to describe both oscillatory and chaotic dynamics. The governing differential equations are:
\[
\begin{aligned}
\frac{dx}{dt} &= -y - z, \\
\frac{dy}{dt} &= x + ay, \\
\frac{dz}{dt} &= b + z(x - c),
\end{aligned}
\]
with $x$, $y$, and $z$ as the system variables, and $a$, $b$, and $c$ as parameters. The system has two fixed points; here, we are considering the fixed point given by 
\[
\begin{aligned}
\begin{pmatrix}
x^* \\
y^* \\
z^*
\end{pmatrix}
= \begin{pmatrix}
\dfrac{c - \sqrt{c^2 - 4ab}}{2} \\\\
\dfrac{-c + \sqrt{c^2 - 4ab}}{2a} \\\\
\dfrac{c - \sqrt{c^2 - 4ab}}{2a}
\end{pmatrix}
\end{aligned}
\]
The dynamics of the system undergo a significant change as $a$ transitions from negative to positive, leading towards oscillatory and chaotic trajectories for values of $c$ considered in the paper.\\

These models serve as fundamental examples of dynamical systems theory, providing insights into the behavior of chemical reactions and chaotic systems, and illustrating basic principles of oscillators within networks.\\

{\subsubsection{Weakly nonlinear analysis}}
\label{subsec:WNL}

{We {begin} by introducing small inhomogeneous perturbations, {\({\bf u}_j\)}, to the uniform equilibrium point, {thereby slightly perturbing the system from its steady state,} {\({\bf x}_j={\bf x}^*_j+{\bf u}_j\)} for \(j=1, \ldots, N\). Substituting this into equations \eqref{eq:systemRD} and performing a Taylor expansion yields the following equation for the time evolution of \({\bf u}_j\):
\begin{equation}\label{eq:equation_u}
\dot{{ \bf u}}_j = {\bf  {J}  u}_j  +  {\bf D} \sum_{k=1}^N {L_{jk}} {\bf u}_k+  \mathcal{M}  { \bf u}_j {\bf u}_j +  \mathcal{N}   {\bf u}_j  {\bf u}_j  {\bf u}_j + \ldots
\end{equation}
where {\({\bf J}\)} is the Jacobian matrix evaluated at the steady state \({\bf x}^*\), and \(\mathcal{M} {\bf u}_j {\bf u}_j\) and \(\mathcal{N} {\bf u}_j {\bf u}_j {\bf u}_j\) represent, respectively, the second and third-order terms. Redefining \(\boldsymbol{\mu}\) above the supercritical Hopf bifurcation as \(\boldsymbol{\mu} = \boldsymbol{\mu}_0 + \epsilon^2 \boldsymbol{\mu}_1\), where \(\boldsymbol{\mu}_1\) is order one and introducing a slow time variable \(\tau= \epsilon^2 t\), the total derivative with respect to the original time \(t\) is then:
\(d/dt \longrightarrow \partial/\partial t + \epsilon^2 \partial/\partial \tau\).
Under the crucially imposed condition \(\kappa=\epsilon^2\), the impact of diffusion is comparable to the deviation from the bifurcation point. In proximity to criticality, matrices \({{\bf J}}\) and operators \(\mathcal{M}\), \(\mathcal{N}\) can be expanded in powers of \(\epsilon^2\): \({{\bf J}} = {{\bf J}}_0 + \epsilon^2 {{\bf J}}_1 + \ldots\), \(\mathcal{M} = \mathcal{M}_0 + \epsilon^2 \mathcal{M}_1 + \ldots\), and 
\(\mathcal{N} = \mathcal{N}_0 + \epsilon^2 \mathcal{N}_1 + \ldots\).
Expanding \(\mathbf{u}\) as a series in terms of both \(t\) and \(\tau\): \(\mathbf{u}_j(t) = \sum_{n=1}^{\infty} \epsilon^n \mathbf{u}_j^{(n)}(t,\tau)\), and substituting this expansion along with the previous ones into Eq. \eqref{eq:equation_u} and grouping terms by order in \(\epsilon\) yields the following set of equations.
\begin{equation}\label{eq:terms}
\left ( \frac{\partial}{\partial t }  {\mathbb{I}} - {{\bf J}}_0 \right ) {\bf u}_j^{(\nu)}={\bf B}_j^{(\nu)}  
\end{equation}
for \(\nu=1,2,3...\) {where $\mathbb{I}$ is the identity matrix} and with \({\bf B}_j^{(\nu)}\) defined according to the superscript $\nu$. 
As shown in \cite{nakao_complex_2014, di_patti_ginzburg-landau_2018}, the solvability condition for the linear system above, as per the Fredholm theorem, is directly satisfied for \(\nu = 1\) and \(\nu = 2\), while it must be explicitly imposed for \(\nu = 3\).
In particular, for \(\nu=1\), one has \({\bf B}_j^{(1)} = 0\), corresponding to the linear problem for the fast variable $t$ with a solution 
\begin{equation}
{\bf u}_j^{(1)}(t, \tau) = W_j(\tau) {\bf U}_0 e^{i \omega_0 t} + \text{c.c.},
\label{eq:linear}
\end{equation}
where \({\bf U}_0\) is the right eigenvector of \({\bf {J}}_0\) corresponding to the eigenvalue \(i \omega_0\). 
Here \(i \omega_0\) is the intrinsic frequency of the identical oscillators and c.c. stands for the complex conjugate of the preceding term. The complex variable \(W_j(\tau)\) represents the amplitude of the perturbation, and its dynamics are assumed to be slow, governed by the slow time scale \(\tau\). This choice is based on the fact that when the system is very close to the threshold and on the unstable side, the growth rate causes a slow increase in amplitude, supporting the consideration of two-time-scale dynamics.
For $\nu=2$ instead, we obtain ${\bf B}_j^{(2)}  =  \mathcal{M}_0 {\bf u}_j^{(1)} {\bf u}_j^{(1)}$. Next, we attempt to represent the solution of the linear system as follows:
\begin{equation*}
{\bf u}_j^{(2)}= W_j^2 {\bf V}_2  e^{2 i \omega_0 t } + \gamma {\bf u}_j^{(1)} + \vert W_j \vert ^2 {\bf V}_0 \, ,
\end{equation*}
for some undetermined constant $\gamma$. Substituting this ansatz into \eqref{eq:terms} and collecting terms independent of $t$, we derive ${\bf V}_0 = -2 {\bf J}_0^{-1} \mathcal{M}_0 {\bf U}_0 {\bf U}^* _0$, where the bar indicates the complex conjugate. Similarly, by grouping the terms proportional to $e^{2 i \omega_0 t }$, we find ${\bf V}_{2} = \left ( 2 i \omega_0 \mathbb{I} -{\bf J}_0 \right ) ^{-1} \mathcal{M}_0 {\bf U}_0 {\bf U}_0$.

The third term in Eq. (\ref{eq:terms}) is given by
\(
{\bf B}_j^{(3)} = \left( {\bf J}_1 - {d}/{d \tau}\mathbb{I} \right) {\bf u}_j^{(1)} + {\bf D} \sum_{k=1}^N L_{jk} {\bf u}_k^{(1)} + 2 \mathcal{M}_0 {\bf u}_j^{(1)} {\bf u}_j^{(2)} + \mathcal{N}_0 {\bf u}_j^{(1)} {\bf u}_j^{(1)} {\bf u}_j^{(1)}.
\)
Now, by explicitly imposing the solvability condition for \(\nu=3\), an amplitude equation for the time evolution of \(W_j(\tau)\) is obtained \cite{nakao_complex_2014, di_patti_ginzburg-landau_2018}, also known as the Complex Ginzburg-Landau (CGL) equation
\begin{equation}\label{eq:GLE}
\frac{d}{d \tau} W_j(\tau) = \sigma W_j -  g \vert W_j \vert ^2 W_j + d \sum_{k=1}^N {L_{jk}} W_k
\end{equation}
where the complex coefficients are given as \(\sigma = \sigma_{Re} + i \sigma_{Im} = ({\bf U}_0^*)^T {{\bf J}}_1 {\bf U}_0\), \(d = d_{Re} + i d_{Im} = ({\bf U}_0^*)^T {\bf D} {\bf U}_0\), and \(g = g_{Re} + i g_{Im} = -(\mathbf{U}_0^*)^\dagger  \left [ 2 \mathcal{M}_0 \mathbf{V}_2 \mathbf{U}^*_0  \right . \left . +2 \mathcal{M}_0   \mathbf{V}_0 \mathbf{U}_0 +  3 \mathcal{N}_0  \mathbf{U}_0   \mathbf{U}_0   \mathbf{U}^*_0   \right ]\) where $\dagger$ indicates the conjugate transpose. Here \({\bf U}_0^*\) is the complex conjugate of \({\bf U}_0\). From here, one can immediately notice the absence of dependence of the CGL coefficients on network topology. The Complex Ginzburg-Landau (CGL) equation serves as a normal form for describing the amplitude of a pattern. In the context of diffusion on one-dimensional continuous support, where {$\bf{L}$} is replaced by $\partial^2/\partial x^2$, the CGL solution manifests as a traveling plane wave, dictating the (slow) spatial modulation of the oscillatory pattern.}\\


\subsubsection{The solvability condition} 
\label{sec:solv}

Fredholm theorem states that the linear equation ${\bf A u}(t) = {\bf b}(t)$ is solvable if $\langle {\bf v}(t), {\bf b}(t) \rangle = 0$ for each vector ${\bf v}(t)$ that satisfies ${\bf A}^* {\bf v}(t) = 0$. Here, ${\bf A}$ denotes a linear operator, with ${\bf u}(t)$ and ${\bf b}(t)$ as complex vectors of equivalent dimensions. 
The adjoint operator ${\bf A}^*$ is defined such that $\langle {\bf A}^* {\bf y}, {\bf x} \rangle = \langle {\bf y}, {\bf A} {\bf x} \rangle$ for any vectors ${\bf x}$ and ${\bf y}$. 
The scalar product is defined as $$\langle {\bf v}(t), {\bf b}(t) \rangle = \int_0^{2 \pi / \omega_0} {\bf v}^{\dagger}(t) {\bf b}(t) dt.$$ 
Referring to equation \eqref{eq:terms}, the Fredholm theorem first necessitates identifying ${\bf v}(t)$ such that $(\partial / \partial t {\mathbb{I}} - {{\bf J}}_0)^* {\bf v}(t) = 0$. 
Considering that ${\bf J}_0$ is a real matrix, integration by parts reveals that $$\left(\frac{\partial \mathbb{I}}{\partial t} - {{\bf J}}_0 \right)^* =-\left(\frac{\partial \mathbb{I}}{\partial t} +{{\bf J}}_0\right)^T.$$ Thus, the equation becomes $-(\partial / \partial t {\mathbb{I}} +{{\bf J}}_0 )^T {\bf v}(t)=0$. 
Following the same discussion in the main text, we seek ${\bf v} (t)$ in the form ${\bf U}_0^* e^{i \omega_0 t}$ for a specific vector ${\bf U}_0^*$. Substituting this ansatz results in $({{\bf J}}_0)^T {\bf U}_0^* = -i \omega_0 {\bf U}_0^*$. 
It is common to normalize ${\bf U}_0^*$ such that $( {\bf U}_0^* )^{\dagger} {\bf U}_0 = 1$, which facilitates calculations.
Defining ${\bf U}_0^*$ allows us to articulate the solvability condition $\langle {\bf U}_0^* e^{i \omega_0 t}, {\bf B}_j^{(\nu)} (t, \tau) \rangle = 0$. The expressions ${\bf B}_j^{(\nu)} (t, \tau)$ prove to be periodic functions with a period of $2 \pi /\omega_0$, they are 
expressed as ${\bf B}_j^{(\nu)} (t, \tau) = \sum_{l=-\infty}^{+\infty} \left({\bf B}_j^{(\nu)} (\tau)\right)_l e^{i l \omega_0 t}$. When this series is multiplied 
by $({\bf U}_0^* e^{i \omega_0 t})^{\dagger}$, it results in periodic functions that, integrated over $2\pi /\omega_0$, result in zero except when $l=1$. 
This leads to $$\langle {\bf U}_0^* e^{i \omega_0 t}, \left({\bf B}_j^{(\nu)} (\tau)\right)_1 e^{i \omega_0 t} \rangle = \int_0^{2 \pi / \omega_0} ({\bf U}_0^*)^{\dagger} \left( {\bf B}_j^{(\nu)} (\tau)\right)_1 dt.$$ Since the integral results in zero only if the integrand is identically zero, the solvability condition simplifies to $({\bf U}_0^*)^{\dagger} \left( {\bf B}_j^{(\nu)} (\tau)\right)_1=0$ $\forall j$ and for all $\nu$.\\


\subsubsection{{Calculation of the} CGL coefficients for the Rössler model}
\label{sec:calcs}

In this section, we provide the calculations for a specific parameter set in the original Rössler model, with the goal of determining the CGL equation coefficients. We begin by evaluating the Jacobian matrix at its fixed point, before perturbation, as given above.
\[
{\mathbf{J}_0} = \begin{pmatrix}
0 & -1 & -1 \\
1 & a & 0 \\
\dfrac{c - \sqrt{c^2 - 4ab}}{2} & 0 & \dfrac{c - \sqrt{c^2 - 4ab}}{2}
\end{pmatrix}
\]
Since carrying on analytical calculations yields lengthy and not understandable formulae, we will proceed numerically considering the parameters as in the main text: \(a=-0.01\), \(b=0.2\) and \(c=30\), which gives
\[ 
{\mathbf{J}_0} = \begin{pmatrix}
0 & -1 & -1 \\
1 & -0.01 & 0 \\
-0.00007 & 0 & -0.00007
\end{pmatrix}
\]
The eigenvalue we consider is \(\lambda_c \approx -0.005 \pm i\), which doesn't have exactly a zero part because \(a\) is not exactly zero. If so, then the fixed point would diverge. However, its real part can be considered negligible at this level. The corresponding right and left eigenvectors are
\[
\mathbf{U}_0 = \begin{pmatrix}
0.00354 + 0.70709i \\
0.70712 \\
-0.00005
\end{pmatrix}
\quad
\mathbf{U}_0^* = 
\begin{pmatrix}
0.00349 + 0.70713i \\
0.70712 - 0.00702i \\
0.70716
\end{pmatrix}
\]

Now let's calculate \({\mathbf{J}_1}\)
\[
{\mathbf{J}_1} = \begin{pmatrix}
0 & 0 & 0 \\
0 & 1 & 0 \\
0.0067 & 0 & 0.0067
\end{pmatrix}
\]
and recalling that \(\sigma = (\mathbf{U}_0^*)^T {\mathbf{J}_1} \mathbf{U}_0 = 0.5+0.0083i\) where clearly \(\sigma_{\text{Re}}>0\). We can use the same eigenvectors to calculate \(d = (\mathbf{U}_0^*)^T \mathbf{D} \mathbf{U}_0=0.2\) recalling that \(D_\phi=D_\psi=D_\chi=0.2\). The remaining coefficient is more involved and is given as 
\[ g = -(\mathbf{U}_0^*)^\dagger  \left [ 2 \mathcal{M}_0 \mathbf{V}_2 \mathbf{U}^*_0  \right . \left . +2 \mathcal{M}_0   \mathbf{V}_0 \mathbf{U}_0 +  3 \mathcal{N}_0  \mathbf{U}_0   \mathbf{U}_0    \mathbf{U}^*_0   \right ] \]
We recall that the second and third order terms here are defined as 
\begin{equation*}
\begin{aligned}
\left ( \mathcal{M} \mathbf{u}_j  \mathbf{u}_j  \right  )_l &=\frac{1}{2 !}  \sum_{m,n=1}^3 \frac{\partial^2 F_l ( \mathbf{x}^*, \boldsymbol{\mu})}{\partial x_n \partial x_m}  ( \mathbf{u}_j )_m    ( \mathbf{u}_j )_n \\
\left ( \mathcal{N}  \mathbf{u}_j  \mathbf{u}_j  \mathbf{u}_j   \right  )_l & =\frac{1}{3 !}  \sum_{m,n,p=1}^3 \frac{\partial^3 F_l ( \mathbf{x}^*, \boldsymbol{\mu})}{\partial x_n \partial x_m \partial x_p}  ( \mathbf{u}_j )_m   ( \mathbf{u}_j )_n ( \mathbf{u}_j )_p 
\end{aligned}
\end{equation*}
Thanks to the dominance of linear terms in the Rössler model, the third order derivatives vanish and thus \(\left ( \mathcal{N}  \mathbf{u}_j  \mathbf{u}_j  \mathbf{u}_j   \right  )_l=0,\,\,\, \forall l\) independently of the choice of the vector \(\mathbf{u}_j\). Due to the same property of the model, we have that only the third equation contributes to the second order term \(\left ( \mathcal{M} \mathbf{u}  \mathbf{v}  \right  )_3=\dfrac{1}{2 !}\left[( \mathbf{u} )_1    ( \mathbf{v} )_3 + ( \mathbf{u} )_3    ( \mathbf{v} )_1\right]\) where to make the notation less confusing and ready to be used in the formula for calculating \(g\), we have considered two different 3-dimensional vectors \(\mathbf{u}, \,  \mathbf{v}\). Now referring to \cite{nakao_complex_2014, di_patti_ginzburg-landau_2018}, we define \(\mathbf{V}_0 = -2  {\mathbf{J}_0^{-1}} \mathcal{M}_0 \mathbf{U}_0 \mathbf{U}^*_0\) and \(\mathbf{V}_{2} = \left ( 2 i \omega_0 {\mathbb{I}} -{\mathbf{J}_0} \right ) ^{-1} \mathcal{M}_0 \mathbf{U}_0 \mathbf{U}_0\) where \(\mathcal{M}_0\) means that the operator as defined above should be evaluated at \(\boldsymbol{\mu}_0\), but in our case this is not relevant since the one partial derivative involved is just a constant. Consequently 
\begin{align*}
    \mathbf{V}_0 &= -2  {\mathbf{J}_0}^{-1}\mathcal{M}_0 \mathbf{U}_0 \mathbf{U}^*_0=\begin{pmatrix}
5.1082\times 10^{-5} \\
5.1082 \times 10^{-3} \\
-5.1082 \times 10^{-3}
\end{pmatrix}\\\\
\mathbf{V}_{2} &= \left ( 2 i \omega_0 {\mathbb{I}} -{\mathbf{J}_0} \right ) ^{-1} \mathcal{M}_0 \mathbf{U}_0 \mathbf{U}_0\\ &=\begin{pmatrix}
-3.8946 \times 10^{-8} - 1.1785 \times 10^{-5}i \\
-5.8922 \times 10^{-6} - 9.98811 \times 10^{-9}i \\
-1.7677 \times 10^{-5} + 8.788 \times 10^{-8}i
\end{pmatrix}
\end{align*}
Proceeding similarly, we finally obtain \(g=1.279 \times 10^{-5} + 0.00255i\). Most importantly, we verified that \(g_{\text{Re}}>0\), indicating that the bifurcation is supercritical. The small value of \(g\) is due to the very weak linearity of the Rössler model and the third entry of the vector \(\mathbf{U}_0\) which is very small.\\


\begin{figure*}
    \centering
    \
    \includegraphics[width=\textwidth]{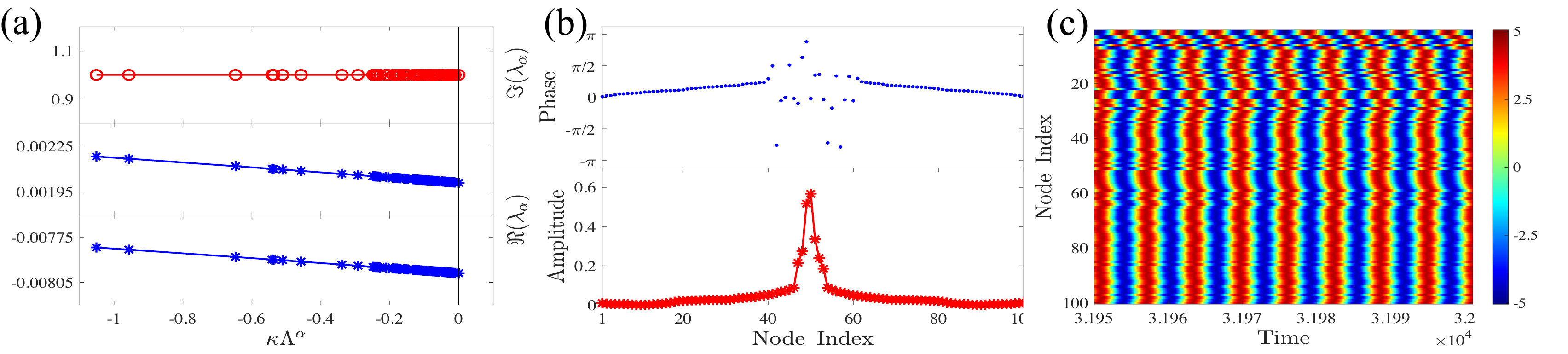}
    \caption{\textbf{Non-chaotic Amplitude-Phase Chimera States in the Rössler Model for the species $\phi$.} (a) Real (blue stars) and imaginary (red circles) parts of the dispersion relation for {$a=-0.01$} (lower) and {$a=0.01$} (middle) respectively. (b) Phase differences (upper) reveal chimera states, with indices scattered for disordered patterns; amplitude differences (lower) show localized disorder, with nodes arranged for a {center-peaked symmetric} amplitude distribution. (c) Pattern evolution at equilibrium with visible amplitude disruptions. The rest of the parameters are $b=0.2$, $c=5.7$, $D_\phi=D_\psi=0$, $D_\chi=0.35$, and $\kappa=0.04$, and the network used is depicted in Fig. \ref{fig:Loc_L}.}
    \label{Ross_fig}
\end{figure*}

\vspace{.25cm}

\subsection{Non-chaotic amplitude mediated chimera patterns}
\label{sec:non_chaos}

Unlike the Rössler model, the Brusselator is characterized by only two variables and, being a two-dimensional system, it cannot produce chaotic behavior according to the Poincaré-Bendixson theorem \cite{Strogatz2015}. Furthermore, as discussed in the main text, the two-dimensionality of the model prevents it from undergoing a Hopf bifurcation in a non-zero spatial domain. Thus, the phase differences from the original limit cycle will always be zero. To observe non-chaotic yet oscillatory behavior in the phase difference, we consider the Rössler equations in a regime where chaos is absent, specifically for a parameter value of $c$ lower than previously analyzed. Figure \ref{Ross_fig} illustrates this scenario (for the species $\phi$), where we particularly emphasize the use of stroboscopic plotting at the same frequency as the original limit cycle. Despite apparent disorder in both amplitude and phase, no quasiperiodicity is present.

\bibliography{nn_bib}

\end{document}